\documentstyle[twoside,fleqn,espcrc2,epsf]{article}
\newcommand{\AmS}{{\protect\the\textfont2
  A\kern-.1667em\lower.5ex\hbox{M}\kern-.125emS}}

\def\lQ{\Lambda_{QCD}}
\newcommand{\nn}{\nonumber}
\newcommand{\be}{\begin{equation}}
\newcommand{\ee}{\end{equation}}
\newcommand{\bea}{\begin{eqnarray}}
\newcommand{\eea}{\end{eqnarray}}

\def\als{\alpha_{s}}
\def\siml{{\
    \lower-1.2pt\vbox{\hbox{\rlap{$<$}\lower6pt\vbox{\hbox{$\sim$}}}}\ }} 

\newcommand{\MS}{\overline{\rm MS}}
\newcommand{\RS}{\rm RS}

\newcommand{\OS}{\rm OS}

\title{Is there a linear potential at short distances?} 

\author{Antonio Pineda\address{Dept. d'Estructura i Constituents de la
  Mat\`eria, U. Barcelona, Diagonal 647, E-08028 Barcelona, Spain} }

\begin{document}

\begin{abstract}
We argue that the lattice data of the static potential can be
explained by perturbation theory up to energies of the order of 1 GeV
once renormalons effects are taken into account.
\end{abstract}

\maketitle

One expects on general grounds the matching coefficients of effective field 
theories of QCD to suffer from renormalon ambiguities. This means that, in principle,
one can not compute these matching coefficients with infinity accuracy
in terms of the (short distance physics) parameters of the underlying
theory. Being more specific, let us consider the effective Lagrangian in some
schematic form:
\be
{\cal L}=\sum_{s=0}^{\infty}{1 \over m^s}c_sO_s
\,,
\ee
where $O_s$ stands for the operators in the effective theory and $c_s$
for the short distance matching coefficients. If we pick up one
matching coefficient $c(\nu/m)$, it would have the
following perturbative expansion in $\als$: 
\be
c(\nu/m)=\bar c+\sum_{n=0}^{\infty}c_n\als^{n+1},
\ee
where $\als$ (in the $\MS$ scheme) is normalized at the scale $\nu$. Its Borel transform
would be
\be
B[c](t)\equiv \sum_{n=0}^{\infty}c_n{t^n \over n!},
\ee
and $c$ is written in terms of its Borel transform as 
\be
c=\bar c+\int\limits_0^\infty\mbox{d} t \,e^{-t/\als}
\,B[c](t)
.
\ee
The ambiguity in the matching coefficient reflects in
poles\footnote{In general, this pole becomes a branch point
singularity but this does
not affect the argumentation.} in the
Borel transform. If we take the one closest to the origin, 
\be
\delta B[c](t) \sim {1 \over a-t},
\ee
where $a$ is a positive number, it sets up the maximal accuracy with which one can obtain the matching
coefficients from a perturbative calculation, which is (roughly) of
the order of 
\be
\label{accpert}
\delta c \sim r_{n^*}\als^{n^*},
\ee
where $n^* \sim {a \over\als}$. 
Moreover, the fact that $a$ is
positive means that, even after Borel resummation, $c$ suffers from a
non-perturbative ambiguity of order 
\be
\label{accnp}
\delta c \sim \left(\lQ\right)^{a\beta_0 \over 2\pi}\,. 
\ee
In order to relate the matching coefficient with
the observable, we use the effective field theory as a tool to provide
the power counting rules in the calculation. Since we are relating one
observable (a renormalon free object) with a matching coefficient
suffering of some renormalon ambiguity, there must be another source
of renormalon ambiguity as to cancel this one. 
The latter comes from the calculation of the matrix element in the
effective theory. If the matrix
element in the effective theory is a non-perturbative object, the
ambiguity of the matching coefficient is of the same size than this
non-perturbative object that we do not know how to calculate
anyhow. Nevertheless, if the matrix element in the effective theory is
a perturbative object, the only way it has to show the renormalon is
in a bad perturbative behavior in the expansion parameters of the
effective theory (what is happening is that the coefficients that
multiply the expansion parameters are not of $O(1)$ due to the
renormalon, i.e. the renormalon is breaking down the assumption of
naturalness implicit in any effective theory). This means that in the
latter situation the observable is less sensitive to
long distance than the matching coefficient itself. Thus, if, for
instance, we wanted later on to get the short distance parameters from
that observable, we are
not doing an optimal job, since we use an intermediate parameter (the
matching coefficient) that can not be obtained with better accuracy
than the ones displayed in Eqs. (\ref{accpert})-(\ref{accnp}), whereas
the observable is less sensitive to
long distances (the same problem also appears if we want to
relate two weakly-sensitive-to-long-distance-physics observables
through a more sensitive to long distance sensitive matching coefficient). 

The renormalon of the
matching coefficients can be {\it spurious} (it is not related to a
{\it real} non-perturbative contribution in the observable) or {\it
real} (it is related to a {\it real} non-perturbative contribution
in the observable). In fact, this distinction depends on the
observable we are considering rather than on the renormalon of the
matching coefficient itself. The point we want to stress is that at the matching
calculation level it makes no sense this distinction. Therefore, there
is no necessity to keep the renormalon ambiguity  in the
matching coefficients. Thus, our proposal is that one should figure out a
matching scheme where the renormalon ambiguity is subtracted from the
matching coefficients. This is the program we will pursue here for the
specific case of effective field theories with heavy quarks and in particular 
for the singlet static potential.

The static potential is the object more accurately studied by
(quenched) lattice simulations. This is due to its relevance in order
to understand the dynamics of QCD. On the one hand, it is a necessary
ingredient in a Schr\"odinger-like formulation of the Heavy
Quarkonium.  On the other hand, a linear growing behavior at long
distance is signalled as a proof of confinement. Moreover, throughout
the last years, lattice simulations \cite{Bali:2000vr,NS}
have improved their predictions at short distances allowing, for the
first time ever, the comparison between perturbation theory and
lattice simulations. Therefore, the static potential provides a unique
place where to test lattice and/or perturbation theory (depending on
the view of the reader), as well as an ideal place where to study the
interplay between perturbative and non-perturbative physics. This is
even more so since the accuracy of the perturbative prediction of the
static potential has also improved significantly recently
\cite{FSP,short,KP1,RG}.

Let us first review the status of the art nowadays.  The prediction
for the perturbative static potential at two loops \cite{FSP}
indicated the failure (non-convergence) of perturbation theory at
amazingly short distances. This failure of perturbation theory is not
solved by the inclusion of the leading logs at three loops computed in
\cite{short,KP1} nor by performing a renormalization group (RG)
improvement of the static potential at next-to-next-to-leading log
(NNLL) \cite{RG}.
On the other hand, it was soon realized that the static potential
suffered of renormalons \cite{Aglietti} and that the leading one (of
$O(\lQ)$) cancelled with the leading renormalon of twice the pole mass
\cite{thesis}. Nevertheless, the prediction of perturbation theory for
the slope of the static potential should not suffer, in principle,
from this renormalon and could be compared with lattice simulations,
which, indeed, only predict the potential up to a 
constant. This comparison was performed in Ref. \cite{Bali}, where
they compared the RG improved predictions
(without including ultrasoft log resummation) of perturbation theory
up to next-to-next-to-leading order with lattice
simulations. They indeed found that the discrepancies with lattice
simulations and the lack of convergence of the perturbative series
remained. They also found that the difference between perturbation
theory and lattice could be parameterized by a linear potential in a
certain range. Therefore, it seemed to support the claims of some
groups \cite{GPZ,S} of the possible existence of a linear potential at
short distances. Such claims contradict the predictions of the
operator product expansion (OPE), which state that the leading
non-perturbative corrections are quadratic in distance at short
distances\footnote{This contradiction is indeed so only if one 
believes that perturbation theory can be applied for the shortest 
distances available in lattice simulations $\sim 1-4$ GeV. We 
thank V.I. Zakharov for stressing this point to us.}.

On the other hand, it has 
been argued \cite{Sumino} that, within a renormalon based picture,
perturbation theory agrees with the phenomenological potentials aimed
to describe Heavy Quarkonium. Moreover, In Ref.  \cite{DESY} 
(see also \cite{Sumino}), it
has been shown that perturbation theory can indeed reproduce the slope
of the static potential given by lattice simulations at short
distances by using the force instead of the potential as the basic
tool without the need to talk about renormalons.

In these proceedings, we would like to review the results of Ref. \cite{staticpot}, 
which tried to clarify further the above
issues and support the renormalon dominance picture by comparing the
potential computed in lattice with the RS static potential (a
renormalon free definition of the potential obtained in
Ref. \cite{RS}).

The energy of two static sources in a singlet configuration reads (in
the on-shell (OS) scheme)
\be
E(r)=2m_{\OS}+\lim_{T\to\infty}{i\over T} \ln \langle W_\Box \rangle
\,.
\ee
In the situation $\lQ \ll 1/r$, it can be computed order by order in
$\als(\nu)$ ($\nu \sim 1/r$) and in the multipole expansion (see
\cite{pNRQCD,RG}) 
\be
E(r)=2m_{\OS}+V_{s,\OS}(r,\nu_{\rm US})+\delta E_{{\rm US}}(r,\nu_{\rm US})
\,.
\ee
$m_{\OS}$ is the pole mass. The static potential reads
($\als\equiv \als(\nu)$) 
\be
\label{potnu}
V_{s,\OS}\equiv V_{s,\OS}(r,\nu_{\rm US})=\sum_{n=0}^\infty V_{n} \als^{n+1},
\ee
where $V_n \equiv V_n(r,\nu,\nu_{\rm US})$. The first three
coefficients $V_0$, $V_1$ and $V_2$ are known \cite{FSP}. The
log-dependence on $\nu$ of $V_3$ can also be obtained by using the
$\nu$-independence of $V_{s,\OS}$. The log-dependence on $\nu_{\rm
US}$ of $V_3$ is also known \cite{short}. Therefore, the only unknown
piece of $V_3$ is a $\nu/\nu_{\rm US}$-independent constant. Its size
has been estimated in Ref. \cite{RS} assuming renormalon
dominance. For $n_f=0$, it reads
\be
\label{V3}
V_{3}(r,1/r,1/r)=1/r \times (-76.1075) 
\,.
\ee
We will use Eq. (\ref{V3}) 
in the following for our estimates of $V_3$.
The static potential is $\nu$-independent. By setting $\nu=1/r$, we
could effectively resum the $\ln{\nu r}$ terms. The RG-improved
expressions would read
\be
\label{potRG}
V_{s,\OS}=\sum_{n=0}^\infty V_{n}(r,1/r,\nu_{\rm US}) \als(1/r)^{n+1},
\ee
and we would have expressions for $n=0,1,2,3$. In the above expression
we have considered $\ln r\nu_{\rm US} \sim 1$. Since $\ln r\nu_{\rm
US} \gg 1$ for some range of the parameters, one could be in the
situation where one also has to resum ultrasoft logs as it has been
done in Ref. \cite{RG}. Nevertheless, explicit calculations show that,
at least for the range of parameters studied in this paper, higher
order ultrasoft logs are subleading even if sizable. However, for
definiteness, we will work with the resummed expression (the physical
picture does not change anyhow), which we will add to
$V_{3}(r,1/r,1/r)$:
\bea
V_{3}(r,1/r,\nu_{\rm US}) &\equiv& V_{3}(r,1/r,1/r)
\\
\nn
&+&
{C_A^3\over
  6\beta_0}\als^3(r^{-1}) \log\left(
\alpha_{s}(r^{-1})\over \alpha_{s}(\nu_{us}) \right)
.
\eea

The RS scheme was defined in Ref. \cite{RS} aiming to eliminate the
renormalons of the matching coefficients appearing in Heavy Quarkonium
calculations. In particular, the leading renormalon of the heavy quark
mass and the static potential. We refer to Ref. \cite{RS} for
details. Here, we just write the relevant formulas needed for our
analysis.

Analogously to the OS scheme, the energy of two static sources in a
singlet configuration reads
\bea
&&
E_s(r)=2m_{\RS}(\nu_f)
\\
\nn
&&
\qquad
+\left(\lim_{T\to\infty}{i\over T} \ln \langle W_\Box
\rangle-2\delta m_{\RS}(\nu_f) \right)
\,,
\eea
where
\bea
&&
\delta m_{\RS}(\nu_f)=\sum_{n=1}^\infty \delta m_{\RS,n}\als^{n+1} 
\\
\nn
&&
=\sum_{n=1}^\infty  N_m\,\nu_f\,\left({\beta_0 \over
2\pi}\right
)^n \als^{n+1}(\nu_f)\,
\\
\nn
&&
\qquad
\times
\sum_{k=0}^\infty c_k{\Gamma(n+1+b-k) \over
\Gamma(1+b-k)}
\,,
\eea
and $m_{\RS}\equiv m_{\OS}-\delta m_{\RS}$. If the beta function were
known to infinity order in perturbation theory, it would be possible
to obtain all the coefficients $b$ and $c_k$ \cite{Beneke2}. In
practice, only $b$ and $c_{0,1,2}$ are known (see
\cite{Beneke2,renormalons,RS}).  For $N_m$ only an approximate
calculation is possible along the line of Ref.
\cite{Lee}. This computation has been done in Ref. \cite{RS} (see also 
\cite{Nm1}). The
result reads (for $n_f=0$)
$$
N_m=0.424413+0.174732+0.0228905=0.622036
,
$$
where each term corresponds to a different power in $u$ (in the Borel
plane) of the calculation (see Ref. \cite{RS} for details). We see a
nice convergence. This number will be the one we will use in the
following. This number should be equal to $-2N_V$, where $N_V$ is the
normalization factor of the renormalon of the static potential. $N_V$
was also approximately computed in Ref. \cite{RS}. For $n_f=0$, it
reads
$$
N_V=-1.33333+0.499433-0.338437=-1.17234
.
$$
In this case the convergence is not so good but we have an alternating
series. In any case, we see that both values appear to be quite
close:
\be
2{2N_m+N_V \over 2N_m-N_V} = 0.059
\,.
\ee
We take this as an approximate indication of the error in the
evaluation of $N_m$.

\begin{figure}[h]
\hspace{-0.1in}
\epsfxsize=2.8in
\centerline{
\put(60,100){$r_0V_{\RS}$}
\put(60,70){$r_0V_{\OS}$(NLO)}
\put(60,50){$r_0V_{\OS}$(NNLO)}
\put(60,25){$r_0V_{\OS}$(NNNLO)}
\epsffile{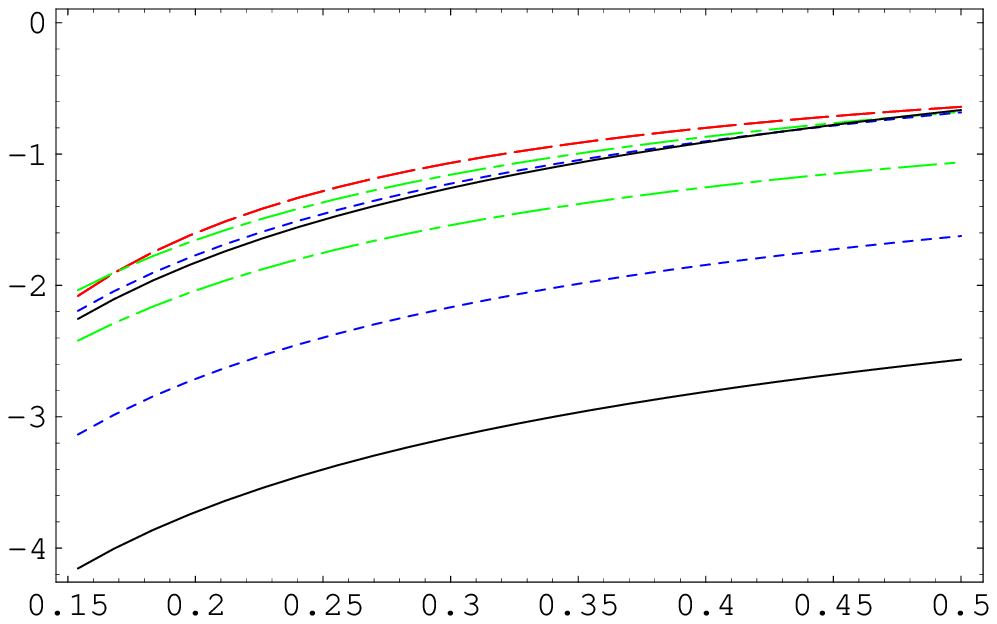}
\put(5,1){$r/r_0$}}
\caption {{\it Plot of $r_0V_{\RS}(r)$ and $r_0V_{\OS}(r)$ 
at tree (dashed line), one-loop
(dash-dotted line), two-loops (dotted line) and three loops (estimate)
plus the leading single ultrasoft log (solid line). For the scale of
$\als(\nu)$, we set $\nu=1/0.15399\,r_0^{-1}$. $\nu_{us}=2.5\,r_0^{-1}$ and
$\nu_f=2.5\,r_0^{-1}$.}}
\label{potRSnu}
\end{figure}

By using $V_{\RS}$, we expect the bad perturbative
behavior that one finds with $V_{\OS}$ due to the renormalon to disappear. 
Let us see that this is indeed so. We consider the OS and RS
potentials at different orders in perturbation theory. We will work with 
$\nu=1/0.15399\,r_0^{-1}$. We display our results in Fig. \ref{potRSnu}. 
We see that we do
not have the gap between different orders in the perturbative
expansion (or it is dramatically reduced) in the RS scheme compared with 
the OS scheme. Although not displayed in these plots, if we had worked with 
$\nu=1/r$, the RS results also converge to the same value.

\begin{figure}[h]
\hspace{-0.1in}
\epsfxsize=2.8in
\centerline{
\put(23,110){$r_0(V_{\RS}(r)-V_{\RS}(r')+E_{latt.}(r'))$}
\epsffile{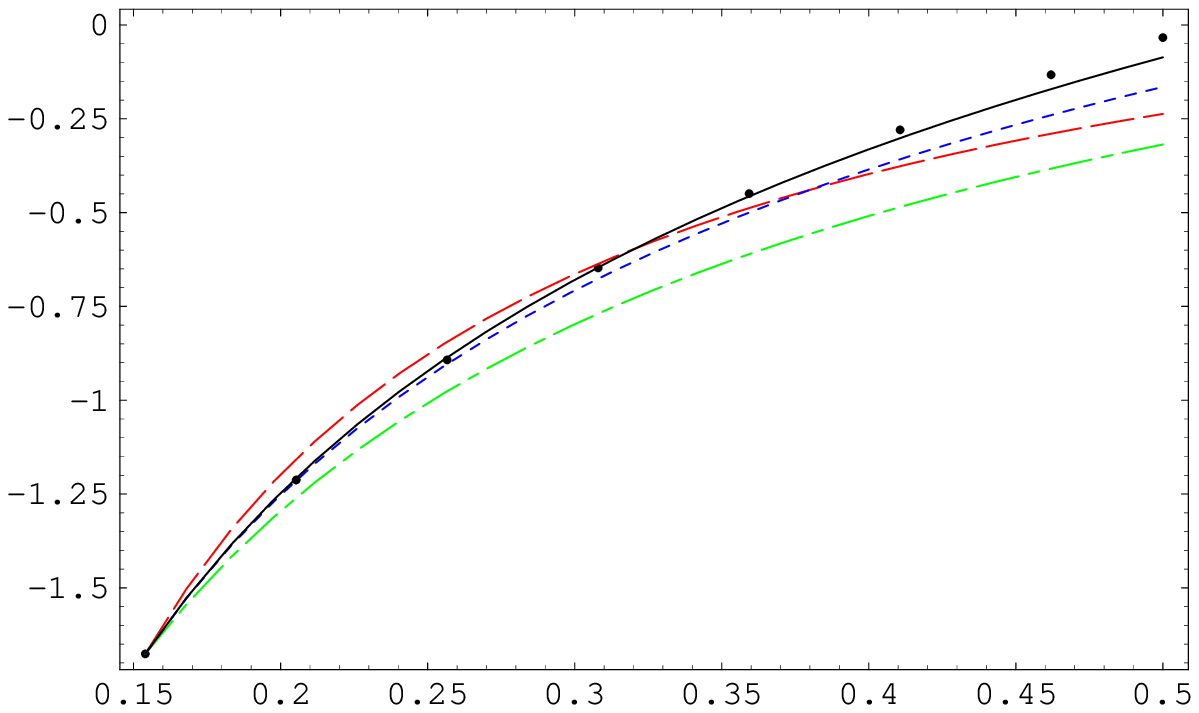}
\put(5,1){$r/r_0$}}
\caption {{\it Plot of $r_0(V_{\RS}(r)-V_{\RS}(r')+E_{latt.}(r'))$ versus
$r$ at tree (dashed line), one-loop (dash-dotted line), two-loops
(dotted line) and three loops (estimate) plus the leading single
ultrasoft log (solid line) compared with the lattice simulations
\cite{NS} $E_{latt.}(r)$. For the scale of $\als(\nu)$, we
set $\nu=1/0.15399\,r_0^{-1}$. $\nu_{us}=2.5\,r_0^{-1}$ and
$r'=0.15399\,r_0$.}} 
\label{potOSlattnu}
\end{figure}

We can now compare with lattice simulations up to a constant in 
Fig. \ref{potOSlattnu}. 
We see that our results are convergent to the
same potential, which, as we can see, corresponds to the lattice
potential.

\begin{figure}[h]
\hspace{-0.1in}
\epsfxsize=2.8in
\centerline{
\put(25,110){$r_0(V_{\RS}(r)-V_{\RS}(r')+E_{latt.}(r'))$}
\epsffile{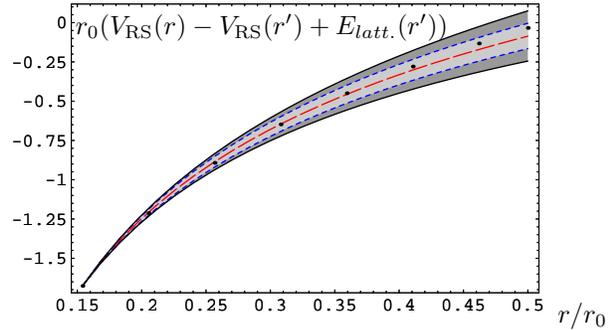}
\put(5,1){$r/r_0$}
}
\caption {{\it Plot of $r_0(V_{\RS}(r)-V_{\RS}(r')+E_{latt.}(r'))$
versus $r$ at three loops (estimate) plus the leading single ultrasoft
log (dashed line) compared with the lattice simulations \cite{NS}
$E_{latt.}(r)$. For the scale of $\als(\nu)$, we set
$\nu=1/0.15399\,r_0^{-1}$. $\nu_{us}=2.5\,r_0^{-1}$ and
$r'=0.15399\,r_0$. The inner and outer band are meant to estimate the
errors in $\Lambda_{\MS}$ and perturbative. For further details see the main
text.}}
\label{errors}
\end{figure}

\begin{figure}[h]
\hspace{-0.1in}
\epsfxsize=2.8in
\centerline{
\put(25,110){$r_0(V_{\RS}(r)-V_{\RS}(r')+E_{latt.}(r'))$}
\epsffile{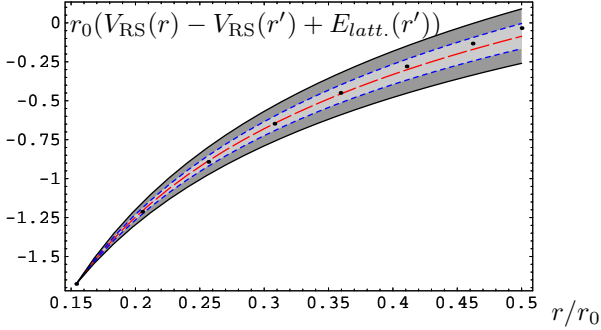}
\put(5,1){$r/r_0$}
}
\caption {{\it Plot of $r_0(V_{\RS}(r)-V_{\RS}(r')+E_{latt.}(r'))$
versus $r$ at three loops (estimate) plus the leading single ultrasoft
log (dashed line) compared with the lattice simulations \cite{NS}
$E_{latt.}(r)$. For the scale of $\als(\nu)$, we set
$\nu=1/0.15399\,r_0^{-1}$. $\nu_{us}=2.5\,r_0^{-1}$ and
$r'=0.15399\,r_0$. The inner and outer band are meant to estimate the
errors in $\Lambda_{\MS}$ and $c_0$. For further details see the main
text.}}
\label{contold}
\end{figure}

We now study possible non-perturbative effects in the static
potential.  Any non-perturbative
effects should be small and compatible with zero since perturbation
theory is able to explain lattice data within errors. We can make this
statement more quantitative by using the lattice data obtained in
Ref. \cite{NS} where the continuum limit has been reached. For these
lattice points the systematic and statistic errors are very small
(smaller than the size of the points). Therefore, the main sources of
uncertainty of our (perturbative) evaluation come from the uncertainty
in the value of $\Lambda_{\MS}$ ($\pm 0.48\,r_0^{-1}$) obtained from
the lattice \cite{Lambda} and from the uncertainty in higher orders in
perturbation theory.  We show our results in Fig.
\ref{errors}.\footnote{We have performed the estimation of the errors
with the results of Fig. \ref{potOSlattnu} ($\nu=constant$), since
they do not introduce errors due to the evaluation of the
renormalon. Nevertheless, a similar conclusion had been achieved if
the analysis had been done with $\nu=1/r$.} 
The inner band reflects the uncertainty in
$\Lambda_{\MS}$ whereas the outer band is meant to estimate the
uncertainty due to higher orders in perturbation theory. We estimate the error 
due to perturbation theory by the difference between the NNLO and NNNLO 
evaluation. This is different from the procedure followed in Ref. \cite{staticpot},
where we allowed $c_0$, the three-loop coefficient of the static potential in 
momentum space (for the specific definition see \cite{staticpot}) to vary by 
by $\pm 146$. We now believe that this may overestimate the size of the 
perturbative errors, as a change of this 
magnitude should also be correlated with the change of the other variables of the 
computation (in particular $N_{V_s}$). Moreover, it would not be consistent with 
the convergent pattern the series shows so far. In any case, the difference is 
not large as one can see by comparing Fig. \ref{errors} with 
Fig. \ref{contold}, for which the 
errors has been obtained with this last method. The difference in the error band, 
not in the central value, of Fig. \ref{contold} with Fig. 9 and, accordingly, 
Fig. 10 of Ref. \cite{staticpot} is due to the fact that we use here the 1-loop 
$\als$ running for the variation, whereas in Ref. \cite{staticpot}, we used the 
three-loop $\als$ running (the use of the three-loop $\als$ running introduces
an incomplete set of subleading logs, therefore we prefer to stick ourselves to the
strict results from the renormalization group method even if the introduction of 
these subleading logs would significantly diminish the errors), plus a missprint in the
formula used for the plot.

\begin{figure}[h]
\hspace{-0.1in}
\epsfxsize=2.8in
\centerline{
\put(25,110){$r_0(V_{\RS}(r)-V_{\RS}(r')$}
\put(25,98){$+E_{latt.}(r')+\sigma (r-r'))$}
\epsffile{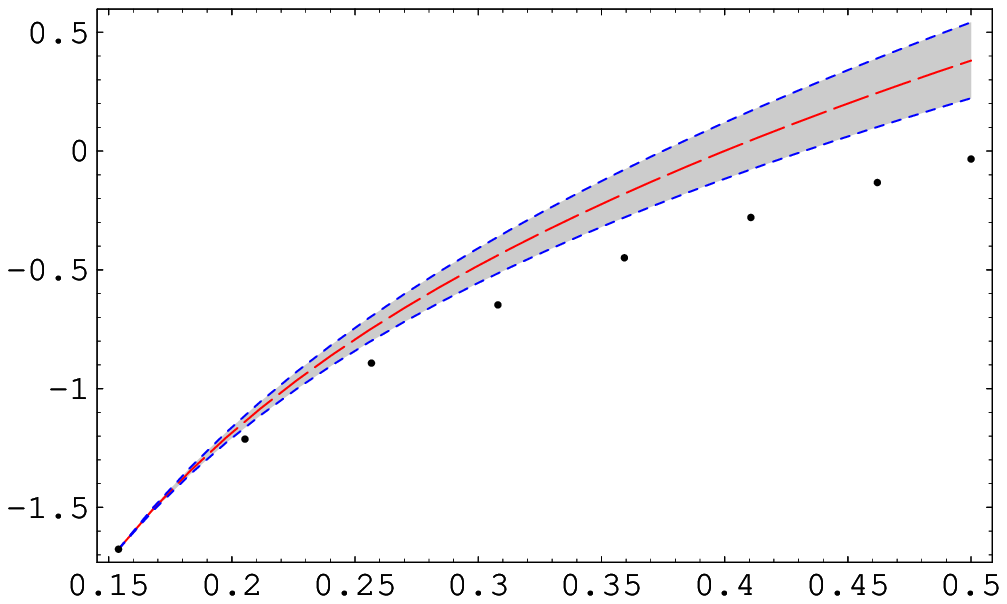}
\put(5,1){$r/r_0$}
}
\caption {{\it Plot of $r_0(V_{\RS}(r)-V_{\RS}(r')+E_{latt.}(r')+\sigma
(r-r'))$ versus $r$ at three loops (estimate) plus the leading single ultrasoft
log compared with the lattice simulations \cite{NS}
$E_{latt.}(r)$. For the scale of $\als(\nu)$, we set
$\nu=1/0.15399\,r_0^{-1}$. $\nu_{us}=2.5\,r_0^{-1}$ and
$r'=0.15399\,r_0$. The dashed band is meant to estimate the
combined error due $\Lambda_{\MS}$ and perturbation theory 
(see Fig. \ref{errors}).}}
\label{linearpot}
\end{figure}

The usual confining potential, $\delta V =\sigma r$, goes with an
slope $\sigma=0.21 {\rm GeV}^2$. In lattice units we take:
$\sigma=1.35\, r_0^{-2}$. The
introduction of a linear potential at short distances with such slope
is not consistent with lattice simulations as we can
see from Fig. \ref{linearpot}. This is even so after the
errors considered in Fig. \ref{errors} have been included. This should
not come as a surprise since this linear potential appears as an
effect of long distance.  Therefore, it follows that the use
of the Cornell potential (with the perturbative static potential
emanated from QCD instead of a pure $1/r$ potential times a fitted
constant) as a phenomenological fit of the static potential introduces
systematic errors if the typical inverse Bohr radius scale of the
heavy quarkonium system to study lye in the short distance regime as
it is the case, for instance, for the $\Upsilon(1S)$.

On the other hand, recently, there have been claims about the possible
existence of a linear potential at short distances
\cite{GPZ,S,Grunberg}. Expected it to be of different physical origin than the
long distance linear potential, it may have a different slope than the
(long distance) static potential discussed above. It would be very
important to discriminate its existence, since such behavior at short
distances is at odds with the OPE.  This is indeed possible in some cases, 
since the short-distance linear potential expected in Ref. \cite{GPZ,S} have an
slope of the order of magnitude of the long-distance confining
potential that we have already ruled out above. Therefore, we can
conclude that no linear potential exists at short distance (with the
present estimates for its slope).

\end{document}